\documentclass[9pt,twocolumn]{article}

\usepackage[
  letterpaper,
  top=0.75in,
  bottom=1in,
  left=0.75in,
  right=0.75in,
  columnsep=0.25in
]{geometry}

\usepackage[utf8]{inputenc}
\usepackage[T1]{fontenc}
\usepackage{microtype}

\usepackage{titlesec}
\titleformat{\section}{\normalfont\bfseries}{\thesection}{0.5em}{}
\titlespacing*{\section}{0pt}{2ex plus 1ex minus .2ex}{0.8ex plus .2ex}
\titleformat{\subsection}[runin]{\normalfont\bfseries\small}{\thesubsection}{0.5em}{}[.]
\titlespacing*{\subsection}{0pt}{1.5ex plus 1ex minus .2ex}{0.5em}

\newenvironment{onecolabstract}{%
  \small
  \begin{center}
    \textbf{Abstract}
  \end{center}
  \begin{quotation}
}{%
  \end{quotation}
}

\usepackage{amsmath,amssymb}

\usepackage{graphicx}
\usepackage{float}
\usepackage[font=footnotesize]{caption}
\usepackage{subcaption}

\usepackage{booktabs}
\usepackage{tabularx}

\usepackage{listings}
\usepackage{xcolor}

\definecolor{codegreen}{rgb}{0,0.6,0}
\definecolor{codegray}{rgb}{0.5,0.5,0.5}
\definecolor{codepurple}{rgb}{0.58,0,0.82}
\definecolor{backcolour}{rgb}{0.97,0.97,0.97}

\lstdefinestyle{pythonstyle}{
    backgroundcolor=\color{backcolour},
    commentstyle=\color{codegreen},
    keywordstyle=\color{blue},
    numberstyle=\tiny\color{codegray},
    stringstyle=\color{codepurple},
    basicstyle=\ttfamily\small,
    breakatwhitespace=false,
    breaklines=true,
    captionpos=b,
    keepspaces=true,
    numbers=left,
    numbersep=5pt,
    showspaces=false,
    showstringspaces=false,
    showtabs=false,
    tabsize=2,
    language=Python
}

\usepackage[colorlinks=true,linkcolor=blue,citecolor=blue,urlcolor=blue]{hyperref}

\usepackage[numbers,sort&compress,super]{natbib}
\newcommand{\datajoint}{\textsc{DataJoint}}
\newcommand{\datajointv}{\textsc{DataJoint~2.0}}
\newcommand{\code}[1]{\texttt{#1}}

\title{\textbf{DataJoint 2.0: A Computational Substrate for \mbox{Agentic Scientific Workflows}}}

\author{\small Dimitri Yatsenko, Thinh T.\ Nguyen --- DataJoint Inc.}

\date{\small February 2026}

\begin{document}

\twocolumn[
\maketitle

\begin{onecolabstract}
Operational rigor determines whether human-agent collaboration succeeds or fails. The most productive AI deployments operate within systems that enable safe iteration: version control, testing, branching, reproducibility. In software engineering, these practices constitute DevOps; scientific data pipelines need an equivalent operational transformation---\emph{SciOps}~\citep{johnson2024sciops}. Yet common approaches to scientific data management make poor substrates for agentic workflows: file-based systems offer flexibility but fragment provenance; task-centric orchestrators manage execution but remain agnostic to data structure; data lakehouses optimize analytical queries but treat computation as external. None provides the unified substrate that agentic scientific workflows require.

\datajointv{} addresses this gap through the \emph{relational workflow model}: tables represent workflow steps, rows represent artifacts, foreign keys prescribe execution order. The schema specifies not only what data exists but how it is derived---a single formal system where data structure, computational dependencies, and integrity constraints are all queryable, enforceable, and machine-readable.

Four technical innovations extend this foundation: (1)~\emph{object-augmented schemas} integrating relational metadata with scalable object storage, (2)~\emph{semantic matching} using attribute lineage to prevent erroneous joins, (3)~an \emph{extensible type system} for domain-specific formats, and (4)~\emph{distributed job coordination} designed for composability with external orchestration. By unifying data structure, data, and computational transformations in a single queryable framework, \datajoint{} creates a computational substrate for SciOps where agents can participate in scientific workflows without risking data corruption.

\end{onecolabstract}
\vspace{1em}
]

\section{Introduction}
\label{sec:intro}

The most successful AI agent deployments in software development share a common pattern: they operate within infrastructure that enforces operational rigor. Version control captures every change. Test-driven development provides continuous validation. Branching allows experimentation without risk to production. CI/CD pipelines automate quality gates. Together, these create a ``ratcheting'' mechanism---progress is captured, mistakes are reversible, and humans steer while automation executes. This is DevOps; the key insight is that \emph{operational infrastructure} determines whether human-agent collaboration succeeds or fails.

Scientific research is undergoing the same agentic transformation. Early AI co-scientist platforms~\citep{huang2025biomni,novelseek2025} emphasize literature synthesis and work with experimental data in small samples supplied to them. But the frontier is shifting toward full integration with scientific pipelines: AI that analyzes results as they are acquired, steers downstream computations, and evolves the analysis pipeline in response to emerging hypotheses---allowing scientists to interact with their entire study at a higher level of abstraction. This is \emph{SciOps}~\citep{johnson2024sciops}: not AI that reads about science, but AI that participates in the scientific workflow. A common belief holds that AI will help make sense of complex data with less need for structure, as long as sufficient metadata is supplied. But operational rigor becomes \emph{more} critical for AI-driven workflows, not less.

Most scientific teams today rely on file-based storage, workflow orchestrators, and ad-hoc scripts---flexible and accessible, but fragmenting provenance across disconnected systems without transactional guarantees, safe experimentation on live pipelines, reversal after errors, or smooth evolution as collaborative teams grow. Relational databases offer the rigor that agentic workflows demand: referential integrity, atomic transactions, declarative queries, machine-readable schemas. Yet they have found limited adoption in science, perceived as rigid, bottleneck-prone, and burdensome to administer. Conventional implementations do little to dispel these concerns: they cannot store large arrays efficiently, describe only \emph{what data exists} rather than \emph{how it is derived}, and disconnect workflows from the data they produce.

\datajoint{} bridges this gap through the \emph{relational workflow model}, which extends relational foundations to represent not merely data structure but the complete scientific workflow (Section~\ref{sec:rwm}). The schema captures workflow specification. Referential integrity prevents corruption. Semantic matching catches erroneous operations. Per-table job coordination enables distributed computation with full provenance. This paper presents \datajointv{}, which further extends these capabilities to meet the demands of agentic computation at scale.

\subsection{Evolution of DataJoint}

\datajoint{} was first released as an open-source MATLAB toolbox in 2011, with the Python implementation following in 2014. The initial publication~\citep{yatsenko2015datajoint} introduced practical implementations for managing scientific data in collaborative settings. Subsequent work formalized the theoretical foundations: entity normalization and a five-operator query algebra~\citep{yatsenko2018datajoint}. DataJoint Elements~\citep{yatsenko2021datajoint} demonstrated modular workflow designs for neurophysiology---reusable schema modules (e.g., for electrophysiology, calcium imaging, pose estimation, behavioral sequencing) that laboratories customize and compose into complete pipelines, enabling consistent data architecture across independent research groups.

\subsection{Contributions}

This paper makes one conceptual and four technical contributions.

\textbf{Conceptual contribution.} We articulate the \emph{relational workflow model} as a distinct paradigm for understanding relational databases---one that adds an operational dimension to classical data modeling. Tables represent workflow steps, rows represent artifacts, and foreign keys prescribe execution order. The schema specifies not only what data exists but how it is derived (Section~\ref{sec:rwm}).

\textbf{Technical contributions.} Building on this foundation, \datajointv{} introduces four innovations for agentic computation:

\begin{enumerate}
    \item \textbf{Object-Augmented Schema (OAS)}: Unified transactional control over relational tuples and object storage (Section~\ref{sec:oas})

    \item \textbf{Semantic Matching}: Lineage-based resolution of matching attributes in binary query operators, preventing erroneous joins on homonymous attributes (Section~\ref{sec:semantic})

    \item \textbf{Extensible Type System}: Pluggable codecs for domain-specific formats (Section~\ref{sec:types})

    \item \textbf{Automated Job Management}: Deterministic per-table distributed computation with provenance tracking (Section~\ref{sec:jobs})
\end{enumerate}

\section{The Relational Workflow Model}
\label{sec:rwm}

The relational data model~\citep{codd1970relational} remains the most rigorous foundation for data management. Its mathematical grounding in predicate calculus and set theory provides precise semantics for data representation and manipulation. Referential integrity, normal forms, and declarative queries enable systems that are consistent, maintainable, and scalable.

The model has historically been interpreted through two conceptual frameworks. Codd's mathematical foundation views tables as logical predicates and rows as true propositions---mathematically rigorous but requiring abstract reasoning disconnected from domain thinking. Chen's Entity-Relationship Model~\citep{chen1976entity} shifted focus to concrete domain modeling with entities, attributes, and relationships---more intuitive, but creating a gap between conceptual diagrams and SQL implementation, and crucially lacking any workflow or computational dimension.

The \emph{relational workflow model}~\citep{yatsenko2018datajoint} introduces a third paradigm: tables represent \emph{workflow steps}, rows represent \emph{workflow artifacts}, and foreign key dependencies prescribe \emph{execution order}. This adds an operational dimension absent from both predecessors.

\begin{table*}[t]
\centering
\begin{tabular}{@{}llll@{}}
\toprule
\textbf{Aspect} & \textbf{Mathematical (Codd)} & \textbf{ER (Chen)} & \textbf{Relational Workflow} \\
\midrule
Core Question & Functional dependencies & Entity types & When/how created \\
Workflow Dimension & Not addressed & Not central & Fundamental \\
Implementation Gap & High & High & None \\
Workflow Support & None & None & Native \\
\bottomrule
\end{tabular}
\caption{Three paradigms for interpreting the relational model.}
\label{tab:paradigms}
\end{table*}

\subsection{Core Principles}

\begin{itemize}
    \item \textbf{Table tiers}: Tables are classified by their data entry mode. \emph{Manual} tables receive direct user entry. \emph{Lookup} tables hold reference data. \emph{Imported} and \emph{Computed} tables define computations via \code{make()} methods. The distinction: Imported tables reach out to data sources outside the DataJoint system (instruments, electronic lab notebooks, external databases), while Computed tables derive their contents entirely from upstream DataJoint tables.

    \item \textbf{Declarative computation}: The \code{make()} method specifies how each entity is derived. This computation logic is declared within the table definition, making it part of the schema itself rather than an external workflow specification.

    \item \textbf{Dependencies as foreign keys}: Foreign keys define computational dependencies, not only referential integrity. The dependency graph is explicit, queryable, and enforced by the database.

    \item \textbf{Master-part relationships}: A workflow step often produces multiple related items that must be created or deleted atomically---for example, detected peaks within a spectrum, or trial events within a session. Master-part relationships declare this transactional grouping directly in the schema: the master table represents the workflow step, while part tables hold the individual items. Insertions and deletions cascade as a unit, enforcing transactional semantics without application code.

    \item \textbf{Entity normalization}: All data is represented as well-formed entity sets with primary keys identifying each entity uniquely. This eliminates redundancy and ensures consistent updates.
\end{itemize}

\subsection{Active vs.\ Passive Schemas}

The key distinction from classical models: traditional schemas are \emph{passive}---containers for data produced by external processes. In the relational workflow model, the schema is \emph{active}---Computed tables declare how their contents are derived, making the schema itself the workflow specification. Schemas are defined as Python classes, and entire pipelines are organized as self-contained code repositories---version-controlled, testable, and deployable using standard software engineering practices.

A useful analogy: electronic spreadsheets unified data and computation in a way that now seems obvious. A spreadsheet naturally contains both cells with values and cells with formulas---it would be unnatural to separate them into distinct systems. Yet this same integration never penetrated relational databases in their 50+ years of history. Classical schemas describe data structure; computation lives elsewhere. The relational workflow model brings to databases what spreadsheets brought to tabular calculation: the recognition that data and the computations that produce it belong together. The analogy has limits: spreadsheets' coupling of data and formulas is also the source of their well-known fragility---hidden dependencies, circular references, ``spreadsheet hell.'' \datajoint{} addresses this through formal schema constraints and explicit dependency declaration rather than ad-hoc cell references; the coupling is governed by relational integrity, not implicit in cell addresses.

This unification has profound implications. Data management (storage, integrity, queries) and workflow specification (dependency resolution, provenance) become a single formal system. The schema captures the entire scientific process: from raw acquisition through derived analyses. Crucially, this enables a clean separation of concerns: scientists define \emph{what} computations derive from \emph{what} data (the workflow specification), while DevOps teams handle \emph{how} computations execute (orchestration)---and can do so uniformly across pipelines without understanding each pipeline's particulars. The open-source framework expresses computational dependencies; orchestration remains a separate operational concern.

\subsection{The Workflow Normalization Principle}

Database normalization decomposes data into tables to eliminate redundancy and ensure that updates propagate consistently. Classical normalization theory~\citep{kent1983simple} achieves this through normal forms based on functional dependencies. Entity normalization~\citep{chen1976entity} asks whether each attribute describes the entity identified by the primary key. Workflow normalization extends these principles with an operational dimension:

\begin{quote}
\textbf{Workflow Normalization Principle}: Every table represents an entity type created at a specific workflow step, and all attributes describe that entity as it exists at that step.
\end{quote}

A Session table contains attributes known when the session is entered (date, experimenter, subject). Analysis parameters determined later belong in Computed tables that depend on Session. This workflow discipline prevents the ``kitchen sink'' tables that plague scientific databases---tables that accumulate attributes from different workflow stages, obscuring provenance and complicating updates.

\subsection{Query Algebra}

\datajoint{} provides a five-operator algebra embedded in Python:

\begin{description}
    \item[Restrict] Filter entities by attribute values or membership in other relations
    \item[Project] Select and rename attributes, compute derived values
    \item[Join] Combine related entities across relations
    \item[Aggregate] Group entities and compute summary statistics
    \item[Union] Combine entity sets with compatible structure
\end{description}

The algebra achieves \emph{algebraic closure}: every operator produces a valid entity set with a well-defined primary key, enabling unlimited composition. For most operators, the result's primary key equals the left operand's primary key; Join is the exception, where the result's primary key depends on functional dependencies between operands. This preservation of entity integrity---every query result is itself a proper entity set with clear identity---distinguishes DataJoint's algebra from SQL, where query results lack both a well-defined primary key and a clear entity type: they are simply a ``bag of rows,'' not a proper entity set.

Queries compose naturally as expressions. The host language's abstraction mechanisms---functions, classes, modules---organize query logic. Scientists work with familiar data structures; the framework handles translation to SQL.

\subsection{Diagram Notation}

\begin{figure*}[h]
    \centering
    \includegraphics[width=1.0\textwidth]{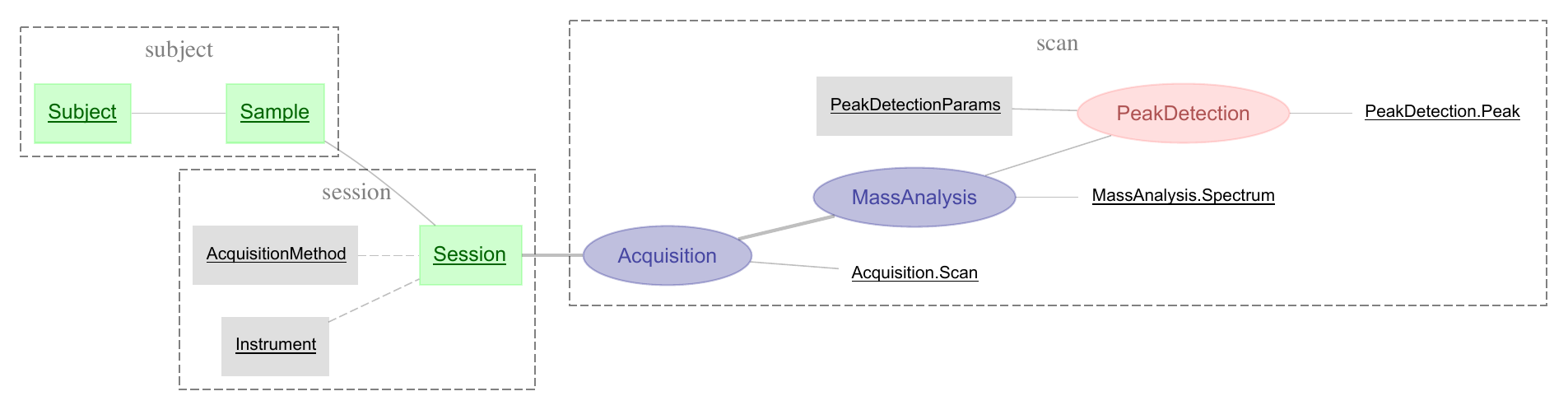}
    \caption{DataJoint diagram of a liquid chromatography--mass spectrometry (LC-MS) data processing pipeline. Green rectangles are manual tables, gray boxes are lookup tables, blue/red ellipses are imported/computed tables. Solid lines indicate identity inheritance; dashed lines indicate references. Part tables (e.g., \texttt{Acquisition.Scan}) appear as plain text. The workflow flows left-to-right: biological samples $\rightarrow$ instrument sessions $\rightarrow$ scan acquisition $\rightarrow$ spectral analysis $\rightarrow$ peak detection. See \texttt{github.com/datajoint/lcms-demo}.}
    \label{fig:lcms-pipeline}
\end{figure*}

\datajoint{} provides auto-generated diagrams that visualize \emph{data workflows}---the directed flow of information through a pipeline (Figure~\ref{fig:lcms-pipeline}). The notation encodes workflow structure at a glance:

\begin{description}
    \item[Table tier] Shape and color distinguish manual tables (green rectangles), lookup tables (gray), imported tables (blue ellipses), and computed tables (red ellipses). Part tables appear as plain text attached to their master.

    \item[Dependency type] Solid lines indicate \emph{identity inheritance}: the child's primary key contains the parent's primary key. Dashed lines indicate references without identity inheritance.

    \item[Directed layout] Source tables appear at the origin; derived results flow toward computed tables. The layout reveals the complete dependency graph and workflow progression.
\end{description}

Dependencies between tables form a directed acyclic graph (DAG); aggregated dependencies between schemas likewise form a DAG. Unlike task DAGs in workflow managers (Nextflow, CWL, Airflow), these are \emph{relational schema} DAGs---they define data structure and relationships, not just execution steps. This structure enables automated reasoning about execution order and parallel scheduling. The diagrams also highlight schema \emph{dimensions}---entities that inject new primary key attributes (e.g., \texttt{Subject}, \texttt{Session}, \texttt{Trial}), rendered with underlined names in the diagram. These primary key sources cascade through dependent tables via identity inheritance. Dimensions define the ``shape'' of data in a pipeline: a table inheriting from both \texttt{Subject} and \texttt{Session} contains one row per subject-session combination. This dimensionality is explicit in the schema and visible in the diagram, enabling agents to reason about data granularity without inspecting table contents.

This notation contrasts with traditional ER diagrams: it shows workflow progression (not just static structure), encodes identity inheritance explicitly (not just cardinality), and is auto-generated from code (always reflecting the actual schema).

\section{Background and Related Work}
\label{sec:background}

\subsection{Related Systems}

SQL remains the universal interface to relational databases, but its limitations are well-known: verbose syntax, poor composability, error-prone semantics (NULL handling, implicit coercion, name-based matching), and results that are ``bags'' rather than proper relations. \datajoint{} retains SQL databases as the storage and transaction engine but provides a different interface: the five-operator algebra described in Section~\ref{sec:rwm}. Against this backdrop, several categories of systems address parts of the scientific data management problem.

\textbf{File-based workflow systems} (Nextflow~\citep{ditommaso2017nextflow}, Snakemake~\citep{molder2021snakemake}, Arvados~\citep{arvados2014}, Cromwell) focus on computational orchestration---scheduling containerized tasks that transform input files into output files. While these systems excel at managing \emph{how} computations run, they treat data as opaque file collections without structural constraints. Metadata is limited to key-value tags; there are no foreign keys, no referential integrity, and no way to query \emph{into} files. Provenance tracks which files produced which files, but not attribute-level lineage. The Common Workflow Language (CWL) provides portability across these systems but inherits their file-centric limitations.

\datajoint{} offers concrete advantages over file-based workflows: \emph{queryable intermediate results} (SQL access to any pipeline stage, not just final outputs), \emph{entity-level provenance} (track which subject or session was affected, not just which workflow ran), \emph{incremental recomputation} (re-run only affected entities when upstream data changes, not entire workflows), and \emph{AI agent integration} (agents can introspect schema structure and query data directly, rather than parsing opaque files and logs).

Decoupled architectures offer genuine advantages that explain their dominance in bioinformatics: independent evolution of data and computation layers (update analysis code without touching the data model), toolchain flexibility (swap compute engines freely---Spark, Dask, GPU clusters---because any tool that reads files works), and natural alignment with organizational boundaries (data engineers, scientists, and DevOps evolve independently). CWL optimized for portability as its top priority and achieved it. \datajoint{} accepts tighter coupling as a deliberate trade-off: framework commitment in exchange for a single formal system where data structure, computational dependencies, and integrity constraints are all queryable, enforceable, and machine-readable. For agentic workflows where agents must reason about data structure, introspect dependencies, and operate safely, the decoupled approach cannot provide what the relational workflow model offers.

Importantly, choosing \datajoint{} does not preclude CWL interoperability. Schema-addressed storage organizes objects in paths that mirror primary key structure (e.g., \code{subject/session/recording.nwb}), matching the directory conventions CWL workflows expect. Migration from CWL to \datajoint{} can be largely AI-assisted; export from \datajoint{} back to CWL is straightforward since the schema already encodes explicit structure. Hybrid operation---where \code{make()} methods invoke CWL tools or CWL workflows populate \datajoint{} tables---enables incremental adoption. There is no platform lock-in.

\textbf{General-purpose workflow orchestrators} (Airflow, Prefect, Flyte, Dagster) provide task-centric execution management: scheduling DAGs of Python functions, managing retries and resources, and monitoring execution state. These systems excel at the operational \emph{when} and \emph{where} of computation but are agnostic to data structure---tasks pass opaque artifacts, and the orchestrator tracks task-level provenance (which DAG ran when) rather than data-level provenance (which inputs produced which outputs). The key distinction: orchestrators schedule \emph{tasks}; \datajoint{} identifies \emph{which data entities need computation} based on schema state. \datajoint{}'s job tables, \code{make()} methods, and foreign key dependency graph map naturally onto orchestration primitives---queues, tasks, and DAGs---making integration with existing orchestration systems straightforward. The open-source library intentionally leaves orchestration (scheduling, resource allocation, retry policies) to external systems; the managed platform integrates these capabilities directly.

\textbf{Data lakehouse systems} (Databricks, Snowflake, Delta Lake, Iceberg) unify structured analytics with scalable object storage using columnar formats optimized for distributed query processing. These systems excel at analytical throughput across massive datasets but treat computation as external to the data model---lineage is captured as metadata rather than enforced by schema constraints, and provenance tracks table-level dependencies rather than attribute-level derivation. \datajoint{} addresses a different part of the data lifecycle: governing how data is derived rather than optimizing how it is queried at scale. In practice, \datajoint{} pipelines produce curated datasets in lakehouse-compatible formats (e.g., Delta tables via the extensible type system and schema-addressed storage on shared object storage), enabling downstream analytical consumption with full upstream provenance.

\textbf{Array databases} (SciDB~\citep{brown2010scidb}, TileDB~\citep{papadopoulos2016tiledb}) optimize for array operations but have weak relational semantics and limited support for heterogeneous scientific metadata.

\textbf{Data versioning systems} (DVC, lakeFS) provide version control for data but lack integrated query languages and computational coordination.

\textbf{AI agent frameworks} (LangChain) enable tool use but provide no data integrity guarantees or transactional safety for database operations.

\section{Object-Augmented Schema Model}
\label{sec:oas}

\subsection{The Scientific Data Challenge}

Scientific datasets combine structured metadata (subjects, sessions, parameters) with large arrays (images, time series, recordings). Relational databases excel at the former: enforcing relationships, maintaining integrity, enabling precise queries. Object stores excel at the latter: efficient storage and access patterns for large binary data. The common workaround---storing file paths in database columns---fragments integrity across two systems: files can be orphaned when records are deleted, references can point to missing objects, and no transactional guarantees span both systems. This is a fundamental, long-standing problem in scientific data management.

The \emph{object-augmented schema} solves this by extending referential integrity to encompass both relational tuples and stored objects under unified transactional control.

\subsection{Proper Representation}

\datajointv{} assigns data to the appropriate storage component based on its nature and access patterns:

\begin{description}
    \item[Relational Database] Structural information, metadata, and small values belong in the relational database, which enforces data integrity through referential constraints and accelerates queries through indexing.

    \item[Object Store] Large scientific objects---arrays, images, recordings---reside in the object store, which affords efficient access patterns: memory mapping, chunking, lazy loading, and parallel access.
\end{description}

Across both storage components, ACID transactional integrity is maintained: the relational database holds lightweight references that ensure atomicity and consistency of all operations.

Within the object store, two addressing schemes serve different needs:

\textbf{Hash-addressed storage} names objects by content hash, enabling automatic deduplication across the entire database. This scheme is appropriate when the same binary object may appear in multiple records---for example, shared configuration files, reference images, or replicated experimental stimuli.

\textbf{Schema-addressed storage} organizes objects in paths that incorporate schema names, table names, and primary key values---enabling direct filesystem navigation without database queries. This naming convention is compatible with file-based workflow systems (CWL, Nextflow, Snakemake), facilitating hybrid operation and migration. The scheme accommodates complex scientific objects potentially composed of multiple parts---chunked arrays (Zarr), hierarchical containers (HDF5), or multi-file datasets.

\subsection{Unified Referential Integrity}

All stored objects participate in referential integrity. Deleting a record cascades to dependent tables \emph{and} removes associated objects. Garbage collection performs an exhaustive check of all references before deleting any object---ensuring that shared hash-addressed objects are removed only when truly unreferenced. Agents can delete data safely; integrity constraints prevent orphans.

\subsection{Transactional Guarantees}

Object operations are transactional: inserts write objects before committing tuples; deletes remove tuples before garbage collecting objects; failures trigger automatic rollback of partial operations.

\section{Semantic Matching}
\label{sec:semantic}

As schemas grow, attributes with identical names may have unrelated meanings. Name-based matching---used by all traditional databases---cannot detect when \code{session\_id} in one table derives from a different source than \code{session\_id} in another, producing spurious joins that silently corrupt results. AI agents face acute risk: they cannot intuit semantic relationships from names alone.

\datajointv{} tracks attribute provenance through foreign key chains. In binary operators (join, restriction, aggregation), same-named attributes match only if they share a common ancestor in the lineage graph; otherwise the query raises an error requiring explicit resolution. Agents receive immediate feedback on query validity---no silent data corruption from semantic mismatches.

\section{Extensible Type System}
\label{sec:types}

\subsection{Three-Layer Architecture}

\begin{description}
    \item[Layer 1: Native Types (Discouraged)] Raw SQL types (\code{TEXT}, \code{TIME}). Backend-specific, not portable. Allowed for legacy compatibility.

    \item[Layer 2: Core Types (Recommended)] Portable types: \code{int64}, \code{float64}, \code{varchar(n)}, \code{datetime}, \code{uuid}, \code{json}. Consistent semantics across MySQL and PostgreSQL.

    \item[Layer 3: Codec Types (Extensible)] Programmable serialization with lazy loading and streaming support. Entry-point based plugin discovery enables domain-specific formats.
\end{description}

\subsection{Lazy References}

Large arrays return lazy references instead of loading data. Metadata (shape, dtype) is accessible without I/O; explicit materialization loads from storage. This enables agents to inspect data properties before committing to expensive operations.

\section{Automated Job Management}
\label{sec:jobs}

\subsection{Deterministic Computation}

Derived data requires computation. \datajoint{}'s job management system is fully deterministic and automated: given a schema with defined dependencies, the system identifies pending work, coordinates distributed workers, handles failures, and tracks provenance---all without requiring intelligent coordination. This determinism is precisely what makes the system reliable for agentic workflows: agents can trigger computations or introduce new data and pipeline branches, relying on the job system to execute derived computations correctly.

\subsection{Job Coordination}

Each computed table has an associated job table tracking reservation status, worker identity, and errors---isolating jobs per computation type with no contention between unrelated pipelines. Workers reserve jobs via optimistic \code{INSERT} (conflicts indicate another worker claimed the job); the \code{populate()} method releases connections during long computations, reacquiring only to insert results.

\subsection{Architectural Boundary}

This job management system provides workflow \emph{coordination}---identifying pending work, reserving jobs, tracking status---but intentionally excludes workflow \emph{orchestration}: scheduling policies, resource allocation, retry strategies, alerting, and monitoring dashboards. This separation is deliberate. \datajoint{} specifies \emph{what} needs to be computed and \emph{why} (because upstream data exists); orchestration systems handle \emph{when} and \emph{where} computations run. The job tables and \code{populate()} mechanism are designed to be composable with external orchestration---Airflow, Prefect, or custom schedulers can read job tables to discover pending work, invoke \code{populate()} calls as tasks, and monitor status through \datajoint{}'s observable state. The managed platform (Section~\ref{sec:implementation}) integrates orchestration directly, removing the need for external orchestration infrastructure for teams that prefer a unified solution.

\section{Suitability for Agentic Workflows}
\label{sec:principles}

Unlike conventional databases, which describe data structure but not computation, and unlike workflow managers, which orchestrate computation but remain agnostic to data structure, \datajoint{} brings together data structure, data itself, and computational transformations into a single queryable framework. This unification is what makes the system a natural substrate for AI agents. Schemas are self-describing: agents can introspect table structure, dependencies, and data state programmatically. Integrity constraints make operations safe by default---invalid joins, type mismatches, and referential violations fail cleanly rather than corrupting data silently. The dependency graph is explicit, enabling agents to reason about execution order without implicit knowledge. Core operations are idempotent, so agents can retry on failure without side effects. And all state---job status, computation progress, errors---is queryable, giving agents the observability they need to monitor and react. Together, these properties mean that agents can participate in scientific workflows with the same transactional guarantees that protect human-initiated operations.

\section{Platform Architecture}
\label{sec:implementation}

A complete \datajoint{} deployment integrates three core components around the open-source library (Figure~\ref{fig:platform}):

\begin{figure*}[h]
    \centering
    \includegraphics[width=0.65\textwidth]{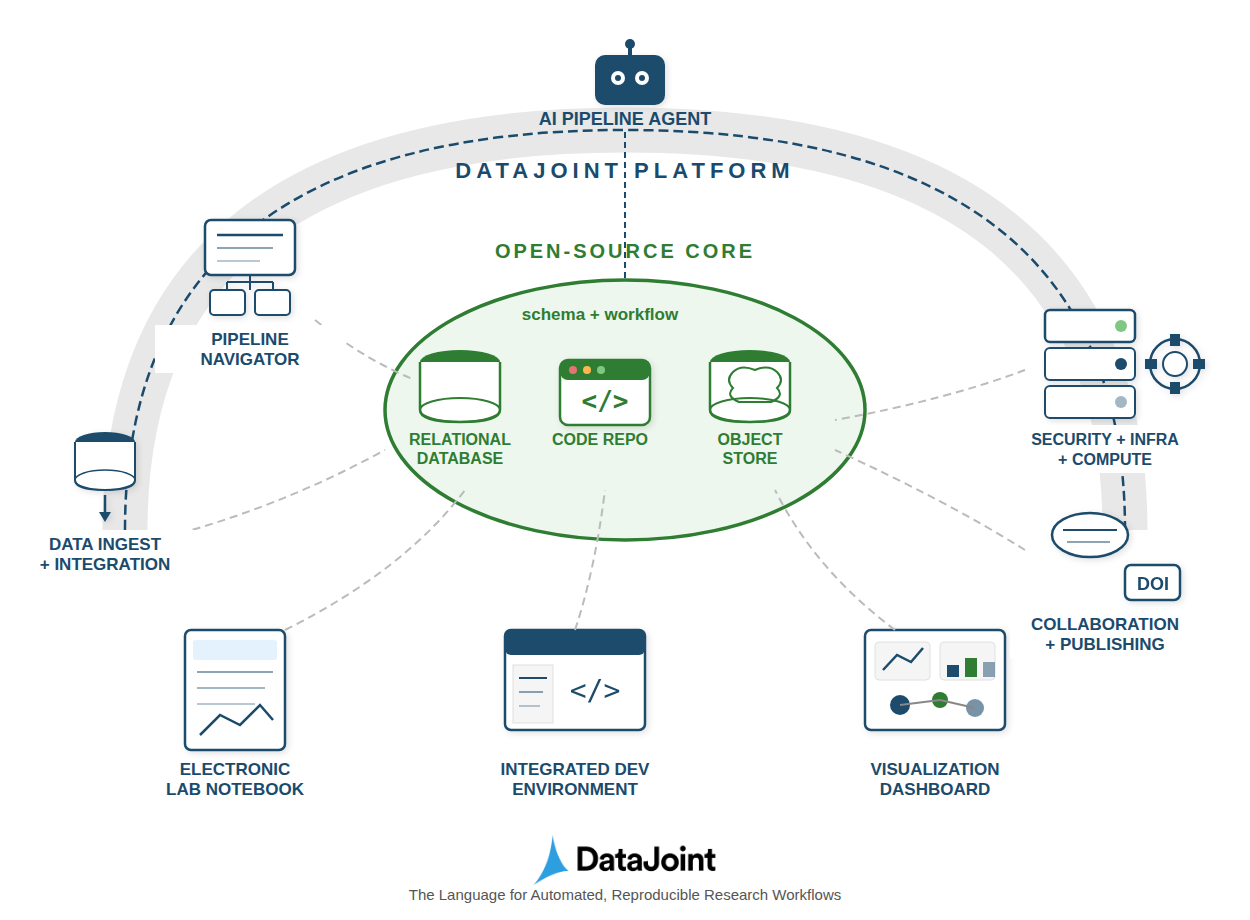}
    \caption{DataJoint platform architecture. The open-source Python library provides the relational workflow model---schema definition, query algebra, and distributed computation. This core integrates with a relational database (system of record), object storage (scalable data), and code repositories (version-controlled pipeline definitions). The managed platform adds infrastructure, observability, and orchestration for production deployments.}
    \label{fig:platform}
\end{figure*}

\subsection{Open-Source Core}

The \datajoint{} Python library~\citep{datajoint-python} implements the relational workflow model: declarative table definitions as Python classes with \code{make()} methods specifying derivation logic, the five-operator query algebra for composable data access, job coordination via \code{populate()} with automatic dependency resolution, and storage abstraction via fsspec supporting local filesystem, S3, GCS, Azure Blob, and custom backends. A pipeline project is a standard Python package---a code repository containing schema definitions, computation logic, and configuration---managed with the same version control, testing, and deployment practices as any software project. SQL backends include MySQL and PostgreSQL. Complete documentation is available online~\citep{datajoint-docs}.

\subsection{Managed Platform}

Production deployments at scale require infrastructure beyond the open-source library: managed databases and object storage, compute orchestration across distributed workers, observability for pipeline health, integration with laboratory information management systems and electronic lab notebooks, and API endpoints enabling AI agents to operate on live pipelines. A managed platform, initially developed with funding from the National Institute of Neurological Disorders and Stroke (NINDS) and now backed by venture capital, addresses these operational requirements for teams across academia, biotech, pharma, and other industries. A growing catalog of existing pipelines provides a fast path to deployment for common experimental modalities. The open-source library and the managed platform share the same pipeline code, allowing teams to choose the deployment model that fits their needs.

\section{Evaluation}
\label{sec:evaluation}

\datajoint{} has supported neuroscience research continuously since 2011. Over 70 peer-reviewed publications describe data pipelines built with the framework, appearing in journals including \emph{Nature}, \emph{Cell}, \emph{Science}, \emph{Nature Neuroscience}, and \emph{Nature Communications}.\footnote{A curated list is maintained at \url{https://docs.datajoint.com/about/publications/}.}

The following projects in life science and precision medicine research illustrate the range of applications: MICrONS~\citep{microns2025} (multimodal neurophysiology and connectomics of mouse visual cortex), Aeon~\citep{aeon2025} (continuous long-term behavioral observation), Spyglass~\citep{spyglass2024} (reproducible hippocampal neuroscience), SCENE~\citep{scene2025} (ecological neuroscience across species), Hussain Shuler Lab~\citep{sutlief2025} (multimodal electrophysiology and behavioral tracking), ORION~\citep{utahorganoids2025} (brain organoid characterization), PosePipe~\citep{cotton2023posepipe,donahue2026posepipe} (clinical markerless motion capture), UCSF Cadwell Lab (multimodal cell-type connectivity), and Harvard Mouse Behavior Core~\citep{weinreb2024kpmoseq} (behavioral syllable extraction via MoSeq). This adoption across independent groups---from millisecond electrophysiology to months-long ethological studies to in vitro systems biology---validates the relational workflow model as a practical foundation for scientific data management.

\section{Discussion}
\label{sec:discussion}

\subsection{Trade-offs}

Additional machinery (lineage tracking, per-table job tables) adds overhead but prevents error classes that are catastrophic at scale. Schema constraints limit ad-hoc operations but ensure data quality over project lifetimes. Production deployments surface practical costs: worker lifecycle management for long-running computations (e.g., GPU model fitting exceeding one hour requires careful timeout configuration), object store tuning for large intermediate artifacts, and connection pool management during multi-hour distributed processing. These operational concerns are inherent to any system that enforces transactional guarantees across distributed computation, but they require expertise beyond the schema definition itself.

The relational workflow model requires learning a uniform framework for data modeling and computation specification. However, teams need to model data and establish computational conventions regardless---\datajoint{} simply provides a consistent way to do so. The payoff is immediate comprehension: researchers who know \datajoint{} can switch between projects and immediately understand how data is structured and computations organized, without deciphering project-specific conventions. The division of labor becomes cleaner: researchers define the pipeline (both data structure and computational dependencies), while DevOps teams handle orchestration in a domain-agnostic manner using standard tools.

Concerns about framework commitment are often overstated. Data lives in standard SQL tables and object stores accessible by any tool. Well-structured pipelines keep analysis logic in independent libraries that \code{make()} methods invoke---these remain portable. What is \datajoint{}-specific is the workflow specification: schema definitions declaring computational dependencies. This parallels any workflow system (Nextflow workflows are Nextflow-specific), but \datajoint{}'s specification lives alongside the data it governs, enabling integrity guarantees and machine-readable semantics for agentic workflows.

An important architectural trade-off: the MySQL/PostgreSQL backends are \emph{row-oriented}, optimized for transactional workloads---inserting rows, reading rows, enforcing referential integrity. This is precisely what \datajoint{}'s workflow model requires. However, row-oriented storage is fundamentally not optimized for analytical workloads: scanning millions of rows across a few columns, large-scale aggregations, or the columnar access patterns that make Spark and lakehouse systems fast. \datajoint{}'s query algebra (Section~\ref{sec:rwm}) is designed for workflow composition (``give me all sessions for subject X with these parameters''), not warehouse-scale analytics (``average firing rate across 50 million neurons grouped by brain region''). \datajoint{} is a pipeline governance and provenance system, not a big-data analytics engine. For large-scale analytics, derived datasets can be exported to lakehouse-compatible formats via the extensible type system and schema-addressed storage, enabling analytical tools to consume governed data with full upstream provenance.

\subsection{Future Directions: Agentic Data Science}

The relational workflow model positions \datajoint{} as a natural substrate for agentic data science. The schema serves as a machine-readable specification that agents can introspect, reason about, and safely modify. Referential integrity and semantic matching prevent data corruption, while job tables and progress tracking provide observable state for monitoring and failure recovery.

The managed platform is actively developing agentic capabilities: natural language interfaces for pipeline exploration, agent-assisted schema development, and autonomous pipeline monitoring. These operational features build on the conceptual foundations described in this paper.

Looking forward, we see the opportunity for collaborative human-agent workflows: interfaces where scientists specify intent at a high level and agents handle implementation details, with the schema serving as the shared contract between human understanding and machine execution. The goal is not to replace scientific judgment but to amplify it---letting researchers focus on experimental design and interpretation while agents handle the engineering of data workflows.

\section{Conclusion}
\label{sec:conclusion}

The relational workflow model offers a new way to understand relational databases---not merely as storage systems but as computational substrates. By interpreting tables as workflow steps and foreign keys as execution dependencies, the schema becomes a complete specification of how data is derived, not just what data exists. This conceptual shift, implicit in DataJoint since its inception, is here articulated as a distinct paradigm alongside Codd's mathematical foundation and Chen's entity-relationship framework.

\datajointv{} extends this foundation for agentic computation: integrating object storage with transactional guarantees, preventing semantic errors through lineage tracking, and enabling distributed computation with full provenance. By unifying data structure, data, and computational transformations in a single queryable framework, \datajoint{} offers a template for building systems that humans and machines can trust.

\appendix
\section{Key Terminology}
\label{sec:terminology}

\begin{description}
    \item[Relational Workflow Model] Tables represent workflow steps, rows represent artifacts, foreign keys prescribe execution order. Adds an operational dimension to Codd's mathematical foundation and Chen's ER model.

    \item[Workflow Normalization] Every table represents an entity type created at a specific workflow step; all attributes describe that entity as it exists at that step.

    \item[Semantic Join] Matches attributes by shared lineage through foreign key chains, not merely shared names. Prevents invalid joins on homonymous but unrelated attributes.

    \item[Object-Augmented Schema (OAS)] Integrates relational tables with object storage under unified transactional control and referential integrity.

    \item[Master-Part Relationship] A workflow step that produces multiple related items as a compositional unit. Insertions and deletions cascade atomically.
\end{description}

\section*{Acknowledgments}

This work was supported by NIH grant U24 NS116470 (DataJoint Pipelines for Neurophysiology), NIH SBIR Direct-to-Phase II grant R44 NS129492 (DataJoint SciOps), and NIH grant R01 NS123849 (ORION). DataJoint's continued development is supported by seed funding from Nina Capital, Inoca Capital Partners, and Capital Factory. We thank DataJoint Inc.\ engineers and scientists Milagros Mar{\'\i}n and Kushal Bakshi for their contributions, as well as open-source consultant Davis Bennett. We also thank the DataJoint open-source community and the many neuroscience laboratories whose adoption and feedback have shaped the system's evolution.

\bibliography{references}

\end{document}